\documentclass[%
 reprint,
 amsmath,amssymb,
 aps,prl
]{revtex4-1}
\usepackage{color}
\usepackage{slashed}
\usepackage{graphicx}% Include  Fig.~files
\usepackage{dcolumn}% Align table columns on decimal point
\usepackage{bm}% bold math
\usepackage{tikz}% bold math
\usepackage[utf8]{inputenc}
\newcommand{\dd}{\,\mathrm{d}}

\newcommand{\A}{\mathcal{A}}
\newcommand{\V}{\mathcal{V}}
\newcommand{\F}{\mathcal{F}}
\newcommand{\HH}{\mathcal{H}}

\newcommand{\aM}{\overline{a}}
\newcommand{\TM}{\overline{T}}

\begin{document}

\preprint{APS/123-QED}

\title{Surface States in Holographic Weyl Semimetals}% Force line breaks with \\

 \author{$\text{Martin Ammon}^1$}
   \email{martin.ammon@uni-jena.de}
  \author{$\text{Amadeo Jim\'enez-Alba}^1$}%
   \email{amadeo.jimenez.alba@uni-jena.de}
   \author{$\text{Markus Heinrich}^{1,2}$}%
   \email[]{markus.heinrich@uni-koeln.de}
 \author{$\text{Sebastian Moeckel}^1$}%
 %\altaffiliation[Also at ]{TPI, FSU Jena.}%Lines break automatically or can be forced with \\
    \email{sebastian.moeckel@uni-jena.de}
\affiliation{%
$\text{\,}^1$Theoretisch-Physikalisches Institut, Friedrich Schiller University Jena, Max-Wien-Platz 1, 07743 Jena, Germany
}%
\affiliation{%
  $\text{\,}^2$Institute for Theoretical Physics, University of Cologne, Z\"ulpicher Stra\ss{}e 77, 50937 Cologne, Germany
  }%


\begin{abstract}
We study the surface states of a strongly coupled Weyl semimetal within holography. By explicit numerical computation of an inhomogeneous holographic Weyl semimetal, we observe the appearance of an electric current restricted to the surface in presence of an electric chemical potential. The integrated current is universal in the sense that it only depends on the topology of the phases showing that the bulk-boundary correspondence holds even at strong coupling. The implications of this result are subtle and may shed some light on anomalous transport at weak coupling.    
\end{abstract}
\maketitle

%\tableofcontents
\section{Introduction}

Weyl semimetals (WSMs) are novel gapless topological states of matter with electronic low-energy excitations behaving as left- and right-handed Weyl fermions \cite{NIELSEN1983389, PhysRevB.83.205101, 2015Sci...349..613X, huang_observation_2015, 2015NatPh..11..748X, lv_observation_2015, yang_weyl_2015}. These quasiparticles are localized around Weyl nodes in the Brillouin zone, points where, in band theory, valence and conduction bands touch at the Fermi energy. The Nielsen-Ninomiya theorem \cite{NIELSEN1983389} states that left- and right-handed Weyl nodes appear in pairs. The existence of these nodes requires either inversion or time reversal symmetry to be broken. In the latter case, Weyl nodes of opposite chirality are separated spatially in the Brillouin zone. Weyl nodes can be characterized as monopoles of Berry flux with charge $\pm 1$ which reflects the chirality of the excitations. As such, Weyl nodes represent topological objects in the Brillouin zone and are stable under most perturbations, including interactions. As in topological insulators, the existence of surface states is guaranteed by topology. Moreover, it has been shown \cite{PhysRevB.83.205101} that the surface states of a WSM form so-called Fermi arcs connecting the projections of the Weyl nodes onto the surface Brillouin zone.

The transport properties of WSMs are tightly bound to the axial anomaly of quantum field theories with Weyl fermions. This leads to anomaly-related phenomena in WSMs such as the anomalous Hall effect \cite{ 2011PhRvB..84g5129Y, 2011PhRvL.107r6806X, 2011PhRvL.107l7205B, 2012PhRvB..86k5133Z, 2013PhRvB..88l5105C}, the chiral magnetic effect \cite{Fukushima:2008xe, Kharzeev:2016mvi} and related effects like the negative magnetoresistance \cite{2013PhRvB..88j4412S, 2014PhRvB..89h5126G}. Furthermore, it has been predicted that lattice deformations couple to the fermionic low-energy excitations with different signs, giving rise to effective axial gauge fields \cite{Cortijo:2015mga, Cortijo:2016yph}. Such lattice deformations naturally arise at the surfaces of a WSM, inducing localised axial magnetic fields in their vicinity. Moreover, it was shown that the Fermi arcs can be understood from this perspective as zeroth Landau levels generated by these fields \cite{Chernodub:2013kya}.

It is important to remark that in the strong coupling regime the description in terms of bands, or even correlation functions, cannot be applied and therefore it is compelling to study to which extent the weak coupling intuitions can be extrapolated. In Refs.~\cite{Landsteiner:2015lsa, Landsteiner:2015pdh}, the authors addressed the question whether the properties of a (homogeneous and infinite) WSM can be found in the strong coupling regime. To this end, they presented a holographic model which exhibits a topological phase transition from a trivial phase to a non-trivial WSM phase. In this letter, we will use this perspective to study surface states of WSMs at strong coupling via holography.

\section{The model}
A (time-reversal breaking) Weyl semimetal can be described by an effective quantum field theory given by \cite{Grushin:2012mt}
\begin{equation}\label{eq:UVmodel}
\mathcal{L}= \bar \psi\left(i\slashed{\partial}-\slashed{\V}-\gamma_5\slashed{\A}+M \right) \psi\, ,
\end{equation}%
where $\V$ is an external $U(1)_V$ gauge field, $\A=\A_i \dd x^i$ is an axial external $U(1)_A$ gauge field (e.g.~induced by strain) and $M$ the fermion mass. The theory is characterized by the axial anomaly
\begin{align}
\partial_\mu j_V^\mu &=0\,,\label{eq:anomaly1}\\ 
\partial_\mu j_A^\mu &=\frac{1}{16\pi^2}\epsilon^{\mu\nu\beta\lambda}\left(\F_{\mu\nu}\F_{\beta\lambda}+3 \HH_{\mu\nu}\HH_{\beta\lambda} \right)   + 2 M \bar \psi \gamma_5 \psi,\label{eq:anomaly2}
\end{align}%
where $\F=\dd \V$, $\HH = \dd \A$ and we have chosen the Bardeen counterterm such that the vector current is non-anomalous. 
 
Let us first consider Weyl nodes homogeneously separated in the $z$ direction, $\A=a\dd z$ with $a=\mathrm{const}$. The model shows two phases, depending on the ratio $a/M$. The global order parameter is the anomalous Hall conductivity (AHC), which can be computed as a one-loop contribution to the effective action \cite{jackiw_when_2000,Grushin:2012mt,2013PhRvL.111b7201V,goswami_axionic_2013}:
\begin{equation}
\label{eq:AHC_weak}
 \sigma_\mathrm{AHE} = \frac{1}{2\pi^2} a_\mathrm{eff} := \frac{1}{2\pi^2}\sqrt{a^2-M^2}\;\Theta(a-M).
\end{equation}
For $a<M$ the AHC vanishes and the spectrum of the theory is gapped with $M_\mathrm{eff}=\sqrt{M^2-a^2}$. For $a>M$ the AHC is finite and the spectrum resembles that of a WSM with gapless modes and node separation $2a_\mathrm{eff}$.

%%%%%%%%%%%%%%%%%%%%%%%%%%%%%%%%%%%%%%%%%
Consider a spatially varying $a(x)$. We expect a contribution to the vector current $\mathbf{j}=\mathbf{j}_V$ by the axial magnetic field $\mathbf{B}_A = -a'(x)\mathbf{e}_y$, similar to the chiral magnetic effect:
\begin{equation}\label{eq:cme}
\mathbf{j}=\frac{\mu}{2\pi^2} \mathbf{B}_A\,,
\end{equation}
where $\mu$ is the (vector) chemical potential. By choosing an appropriate $a(x)$, we can model an interface between the two phases. Then, $\mathbf{B}_A$ is localised at the interface and Eq.~\eqref{eq:cme} gives rise to a surface current. As shown in Ref.~\cite{Chernodub:2013kya}, Fermi arc states can be reinterpreted as zeroth Landau level states coming from $\mathbf{B}_A$ and the above contribution to $\mathbf{j}_V$ is a direct signature of these surface states. Analogous situations were recently studied at weak coupling using a tight binding model in  Refs.~\cite{2016arXiv160704268G, Cortijo:2016wnf}.

Strictly speaking, Eq.~\eqref{eq:cme} is only valid for $M=0$ and a homogeneous $\mathbf{B}_A$, which is the case studied in Ref.~\cite{2016arXiv160704268G}. Nevertheless, inhomogeneous $\mathbf{B}_A$ naturally arise at WSM surfaces. Then, we need higher order correction terms which are not completely determined by the axial anomaly. However, in holography, one can explicitly compute the current taking into account these corrections.

%%%%%%%%%%%%%%%%%%%%%%%%%%%%%%%%%%%%%%%%%
\subsection{Holographic model}
The holographic model in $\text{AdS}_5$, first introduced in Ref.~\cite{Landsteiner:2015lsa, Landsteiner:2015pdh}, consists of two Abelian gauge fields, $V=V_a \dd x^a$ (``vector'') and $A= A_a \dd x^a$ (``axial'') with curvatures $F=\dd A$, $H=\dd V$ in the bulk coupled via a Chern-Simons term as well as a scalar field $\Phi$ charged under the axial field:
\begin{align}
\mathcal{L}=&-\frac{1}{4}H^{ab}H_{ab}-\frac{1}{4}F^{ab}F_{ab}+(D_a\Phi)^*(D^a\Phi)-U(\Phi)\nonumber\\
&+\frac{\alpha}{3}\epsilon^{abcde}A_a\left(F_{bc}F_{de}+3 H_{bc}H_{de}    \right)\,,
\end{align} %%
where $U(\Phi)=m^2|\Phi|^2$ and $D_a\equiv \partial_a-i q A_a$ \footnote{See appendix for a discussion on the choice of parameters. See also Ref.~\cite{Copetti:2016ewq} for further considerations.}. The form of the Chern-Simons term is chosen such that the one-point functions for the divergences of the dual consistent currents match the form of Eq.~\eqref{eq:anomaly1} and Eq.~\eqref{eq:anomaly2}. In the following, $\rho$ is the holographic coordinate with conformal boundary at $\rho=0$.

\paragraph{Anomalous Hall effect.} In order to reproduce the previously discussed phase transition and its features, the boundary values of the fields at $\rho=0$ are identified with $a$ and $M$ as follows:
\begin{align}
A_z(0) &= a, & \varphi(0) &= M \,,
\end{align}
where $\Phi=\rho\varphi$ (see Eqs.~(12-15) in the appendix). Indeed, at zero temperature, this holographic model exhibits a topological phase transition depending on $\aM=a/M$ from a topologically trivial phase for $\aM<\aM_c$ to a topologically non-trivial WSM phase for $\aM>\aM_c$ \cite{Landsteiner:2015pdh}.

Moreover, at finite temperature, this topological phase transition becomes a crossover and gives rise to a critical region in the phase diagram \cite{Landsteiner:2015lsa,Landsteiner:2015pdh}. Eventually, at high enough temperatures, the character of the phase transition is lost.

In the following, we introduce an event horizon at $\rho=1$ to study the holographic model at finite temperature. This allows us later on to study the interface effects not only as a function of the phases but also of the temperature.

The order parameter of the topological phase transition is the anomalous Hall effect. It was shown in  Ref.~\cite{Landsteiner:2015lsa, Landsteiner:2015pdh} that the expression for the anomalous Hall conductivity (AHC) in a homogenous system is 
\begin{equation}
\label{eq:AHC_hol}
\sigma_\mathrm{AHE}=8\alpha A_z\big|_{\rho=1}\,.
\end{equation}
Note that the holographic result is very similar to Eq.~\eqref{eq:AHC_weak}, provided that we identify $\alpha=1/16\pi^2$. However, the effective coupling $a_\mathrm{eff}=A_z|_{\rho=1}$ deviates in general from the weak coupling result  \cite{Landsteiner:2015lsa, Landsteiner:2015pdh}. Concretely, it was found that the onset of the order parameter happens at $\aM_c\approx 1.4$, in disagreement with the weak coupling expectation $\aM_c=1$. 

\begin{figure}[t!] 
\centering
\includegraphics[width=\linewidth]{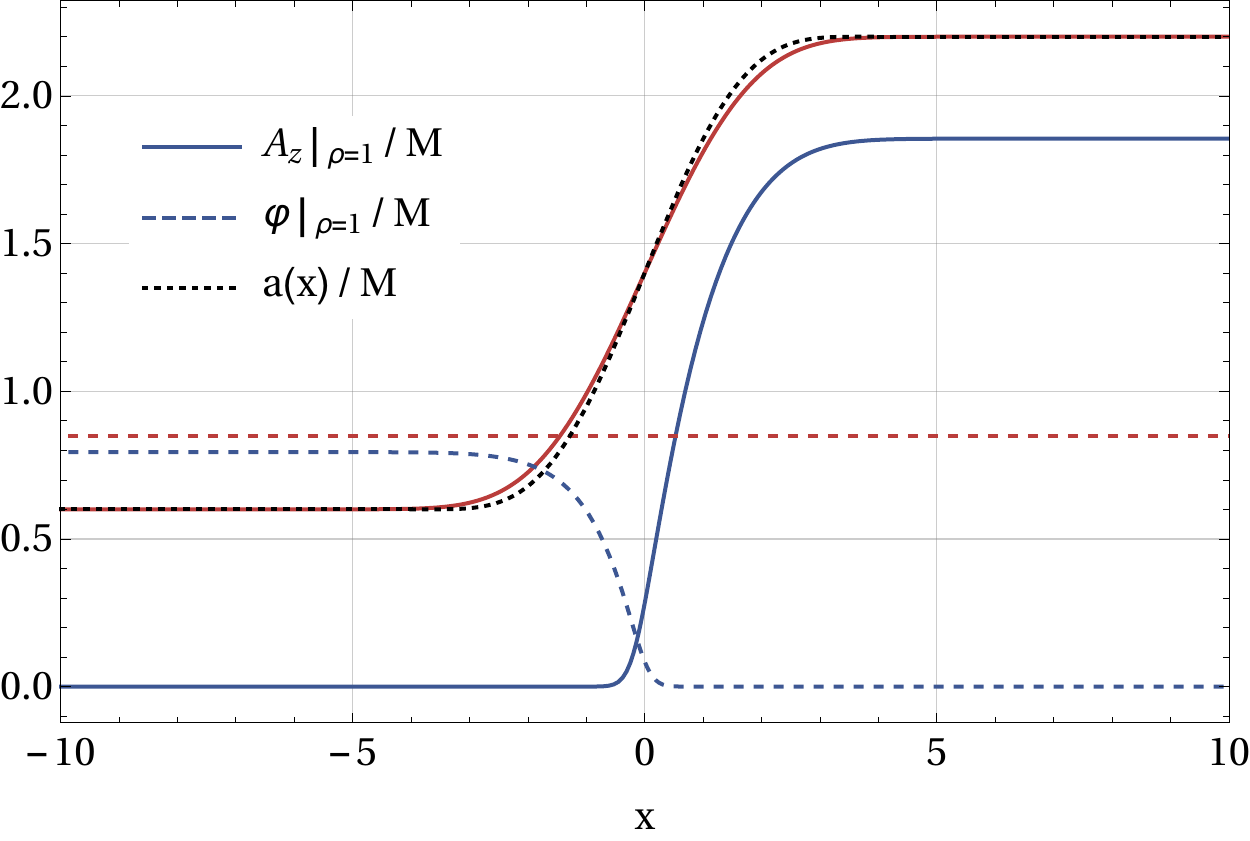}
\caption{Horizon values of scalar and axial gauge field (dashed, solid) along the sample for temperatures $\pi \TM := \pi T/M=100,1/16$ (red, blue). The black dotted line corresponds to the profile of $a(x)/M$.}
\label{fig:AHC}
\end{figure}

\paragraph{Interfaces.} We realize an interface between different phases by implementing step-like profiles for the time-reversal breaking parameter $a(x)$ while setting a homogeneous mass $M$. For numerical reasons, we choose the profile to be the following function:
\begin{equation}
 \aM(x) = \begin{cases}
               \aM_L & \text{for } x < -l \\
               p(x) & \text{for } x \in [-l,l] \\
               \aM_R & \text{for } x > l
              \end{cases}.
\end{equation}
Here, $p(x)$ is an appropriate $C^2$ spline interpolation, see the black dotted line in  Fig.~\ref{fig:AHC} for an explicit example. 

In order to induce a surface current, i.e. to populate the surface states, we need to introduce a vector chemical potential $\mu$. In holography this requires a non-trivial profile for $V_t=V_t(x,\rho)$, which in addition implies a non-trivial $V_y=V_y(x,\rho)$. Since we do not want to explicitly source the surface current, the appropriate boundary conditions read   
\begin{align}\label{eq:bc}
A_z(x,0) &= a(x), & \varphi(x,0) = M, \\
V_t(x,0) &= \mu, & V_y(x,0) = 0,
\end{align}
accompanied by $V_t(x,1)=0$. The vector current can then be computed from the holographic action yielding
\begin{align}
  j^y(x) = \lim_{\rho\rightarrow 0} \sqrt{-g} \left( F^{\rho y} - 4 \, \alpha \, \varepsilon^{\rho y bcd} A_b F_{cd} \right) = \frac{\partial^2V_y}{\partial\rho^2}\bigg|_{\rho=0}.
\end{align}
Note that due to the Minkowski metric on the boundary, $j^y=j_y$. We obtained the following results by solving numerically the equations of motion for a fixed AdS Schwarzschild background \footnote{See appendix}.

%%%%%%%%%%%%%%%%%%%%%%%%%%%%%%%%%%%%%%%%%%%%%%
%%%%%%%%%%%%%%%%%%%%%%%%%%%%%%%%%%%%%%%%%%%%%%
 Let us remark at this point that the holographic direction can be interpreted as an energy scale for the renormalization group (RG) flow. In this context, the horizon values of the fields $A_z$ and $\varphi$ can be interpreted as the IR values of the UV couplings $a$ and $M$. In  Fig.~\ref{fig:AHC}, we show these values along the $x$ direction at high and low temperatures for $\aM_L<\aM_c$ and $\aM_R>\aM_c$. Note that away from the interface the system is homogeneous and thus Eq.~\eqref{eq:AHC_hol} applies. As expected, at low temperatures, $a_\mathrm{eff}$ vanishes in the trivial, left phase, implying that the AHC is zero. Moreover, the non-vanishing scalar field can be interpreted as an effective non-zero mass. In the non-trivial WSM phase on the right, where the driving parameter $\aM$ is above the critical value everywhere, the effective mass vanishes and the phase is characterized by $a_\mathrm{eff}>0$ and a non-vanishing AHC. The fact that we do not observe a sharp transition is a result of the finite extent of the interface and the non-zero temperature. In order words, we are probing the critical region in the phase diagram instead of the critical point of the quantum phase transition itself. A similar behaviour was observed in Refs.~ \cite{Landsteiner:2015lsa, Landsteiner:2015pdh}. Let us remark that these profiles depend not only on the temperature, but on the concrete profile of $\aM(x)$, too.

\section{Results}

\begin{figure} 
\centering
\includegraphics[width=\linewidth]{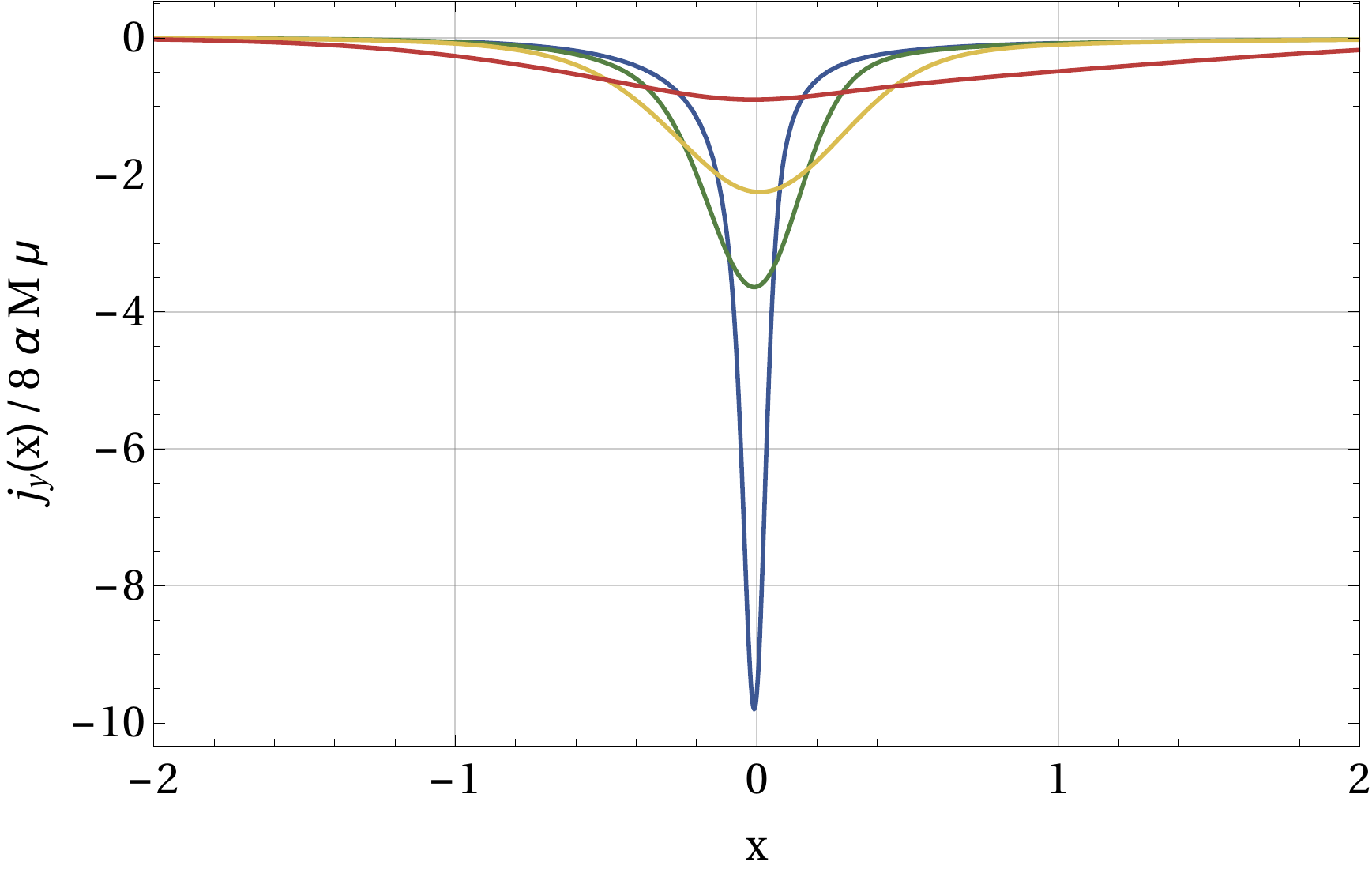}
\caption{Current profile along the $x$ direction for different interface widths $l=0.1,0.5,1,4$ (blue, green, yellow, red) and low temperature $\pi \TM=1/16$.}
\label{fig:current_profile}
\end{figure}

\paragraph{Current profiles.}
Let us first study the profiles of the computed currents. Note that there is no scaling symmetry in $x$ direction which allows us to choose an arbitrary, finite value for the interface width $l$. This is due to the choice $\rho=1$ for the event horizon and thus all lengths are measured in units of inverse temperature. Figure \ref{fig:current_profile} shows the electric current $j_y$ along the sample for the profile in Fig.~\ref{fig:AHC} and different choices of $l$ at low temperature. We observe that as $l$ decreases, the current profile becomes more localised at the interface $x=0$. Thus, it is justified to call $j_y$ a surface current. However, as we checked explicitly for the whole parameter space considered in the next section, $J_y$, the integral of the current along the $x$ direction, is independent of $l$. Consequently, the amplitude of the current has to increase with decreasing $l$. Let us further remark that $j_y$ deviates from Eq.~\eqref{eq:cme}, even after replacing $a'$ with $a'_\mathrm{eff}$ as expected from a non-vanishing mass $M$.

%%%%%%%%%%%%%%%%%%%%%%%%%%%%%%%%%%%%%%%%%%%%%%
%%%%%%%%%%%%%%%%%%%%%%%%%%%%%%%%%%%%%%%%%%%%%%

\begin{figure}
\centering
\includegraphics[width=\linewidth]{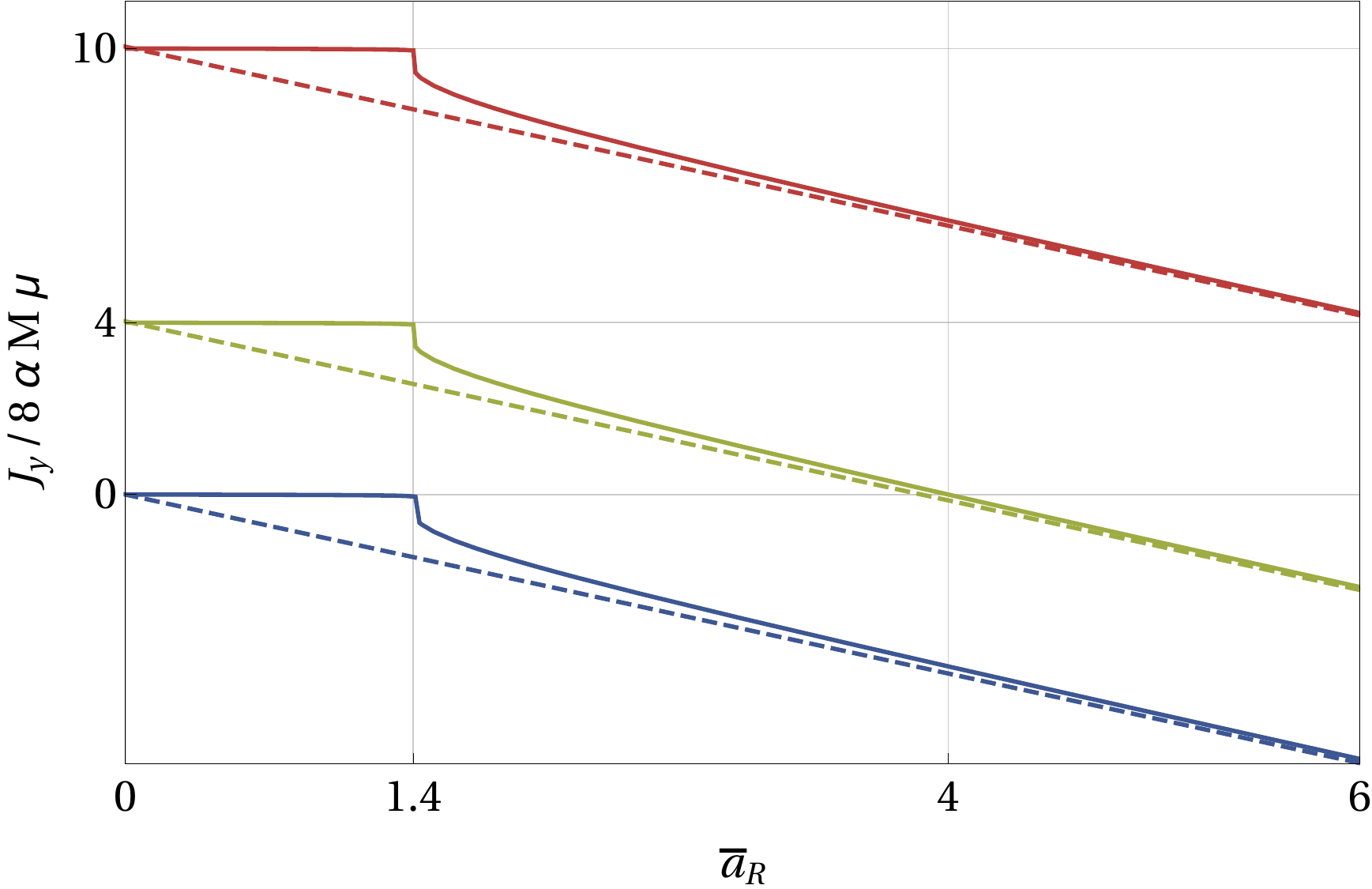}

\caption{Integrated current as a function of $\aM_R$ for $\aM_L<\aM_c$ (blue) and $\aM_c<\aM_L \in \{4, 10\}$ (green, red). Solid lines correspond to low temperature $\pi \TM=1/16$ and dashed lines to high temperature $\pi \TM=1$.}
\label{fig:intcurrent_slices}
\end{figure}

\paragraph{Integrated current.}
In figure \ref{fig:intcurrent_slices}, we show the integrated current $J_y$ for several fixed values of $\aM_L$ as a function of $\aM_R$ and for different values of the temperature/mass ratio $\TM=T/M$. We will refer to this quantity as the temperature from now on. 

For high temperatures $\TM$, as shown by the dashed lines in Fig.~\ref{fig:intcurrent_slices}, we recover the integrated version of Eq.~\eqref{eq:cme} with $\mathbf{B}_A=\nabla \times \mathbf{\mathcal{A}}=-a'(x) \mathbf{e_{y}}$:
\begin{equation}\label{eq:uvintegratedcurrent}
J_y=-\frac{\mu}{2\pi^2}\int_{-\infty}^{\infty} dx \,a'(x)= \frac{\mu}{2\pi^2}(a_L-a_R)\,.
\end{equation}
This result is in line with our expectation for high temperatures, namely that the mass becomes negligible and hence the two originally topologically distinct phases become indistinguishable. Thus, the interface is not a topological phase boundary anymore and the dependence of $J_y$ on $a_{L/R}$ is independent of their concrete values.

For low temperatures, however, the behaviour is different. Let us first consider the situation where the material to the left of the interface is topologically trivial. As shown by the solid blue line in Fig.~\ref{fig:intcurrent_slices}, no current is generated until the r.h.s. of the sample undergoes the phase transition. For $\aM_R$ beyond the critical value $\aM_c$, a finite current is observable which depends nonlinearly on $\aM_R$. Although it seems that the curve has a kink at $\aM_c$, it is in fact smooth due to the finite temperature of the system. 

Moreover, these results do not depend on the concrete value of $\aM_L$ as long as $\aM_L<\aM_c$. In other words, the current is insensitive to the details of the trivial phase. In particular, no current arises at the interface of two trivial phases.

%%%%%%%%%%%%%%%%%%%%%%%%%%%%%%%%%%%%%%%%%%%%%%
%%%%%%%%%%%%%%%%%%%%%%%%%%%%%%%%%%%%%%%%%%%%%%

Let us now consider the case of a non-trivial l.h.s, $\aM_L>\aM_c$, as shown by the solid green and red lines in  Fig.~\ref{fig:intcurrent_slices}. We observe that the difference to the previous situation is simply an overall shift of the integrated current depending on $\aM_L$. This can be understood as follows: The plateau for $\aM_R<\aM_c$ again arises from the insensitivity of the current to the parameter of the trivial phase. However, the plateau value is now non-zero since the l.h.s.~of the sample is non-trivial. Once the r.h.s. undergoes the phase transition, both sides are in the non-trivial phase and the current shows the same functional dependence on $\aM_R$ as before. Note that the shift guarantees that the current vanishes exactly for $\aM_L=\aM_R$ (shown by the solid green line in Fig.~\ref{fig:intcurrent_slices}).

\begin{figure} 
\centering
\includegraphics[width=\linewidth]{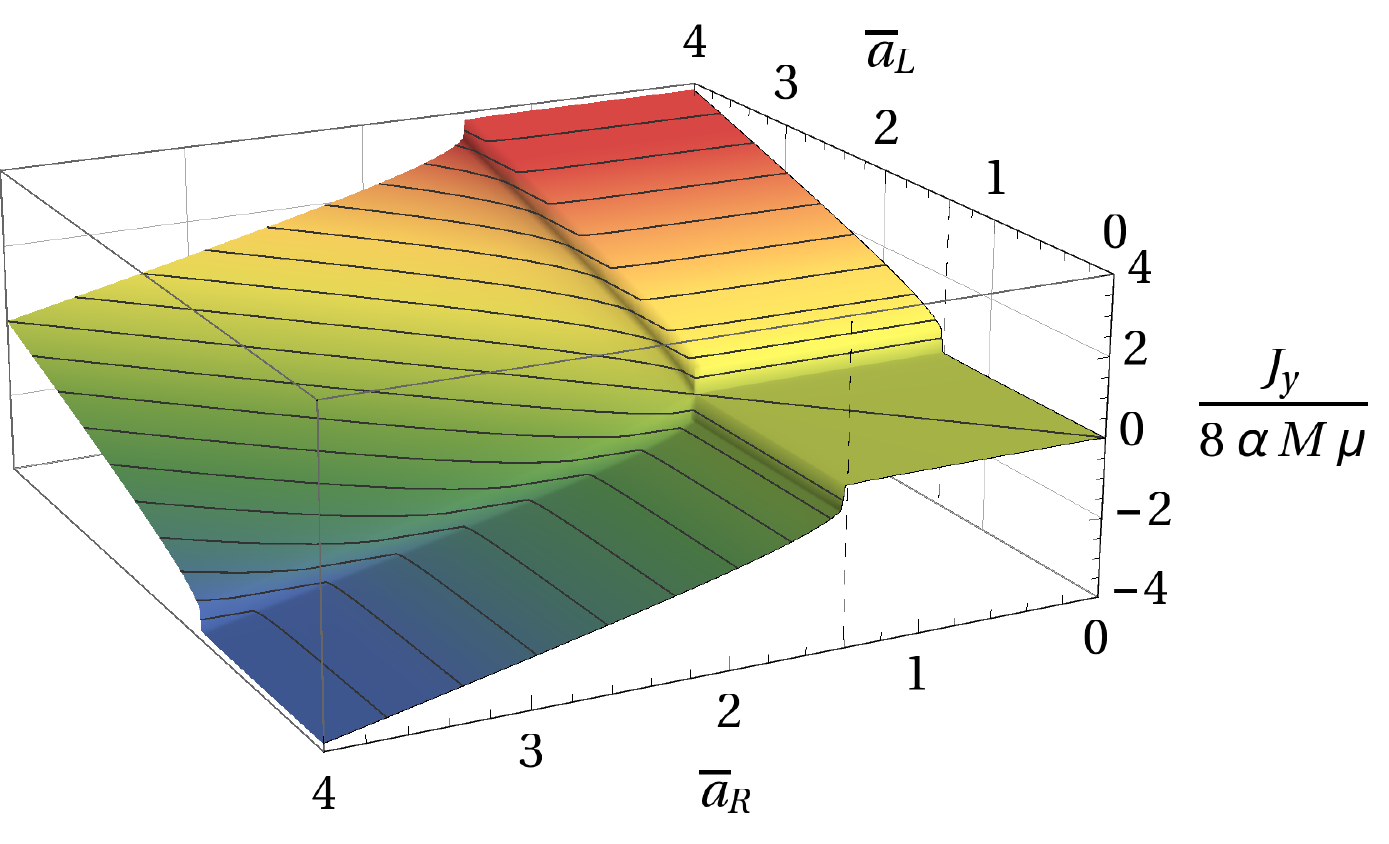}
\caption{Integrated current as a function of $\aM_R$ and $\aM_L$ at low temperature $\pi \TM=1/16$. Contour lines shown in black.}
\label{fig:intcurrent_3D}
\end{figure}

Having discussed the current as a function of $\aM_R$, we show the full dependence of the integrated current at low temperatures in Fig.~\ref{fig:intcurrent_3D}. As in Fig.~\ref{fig:intcurrent_slices}, we find a plateau region for $\aM_{L/R}<\aM_c$. Furthermore, a fundamental symmetry of the system is manifest in this figure, namely that the current is odd under the left-right exchange $\aM_L \leftrightarrow \aM_R$ which reflects its chiral nature \footnote{This property can be shown from the transformation of the equations of motion under the reflection $x\mapsto -x$.}.

%%%%%%%%%%%%%%%%%%%%%%%%%%%%%%%%%%%%%%%%%%%%%%
%%%%%%%%%%%%%%%%%%%%%%%%%%%%%%%%%%%%%%%%%%%%%%

The previous discussion implies that the current is of the form of Eq.~\eqref{eq:uvintegratedcurrent} where $a_{L/R}$ is replaced by an appropriate function $\mathcal{F}(a_{L/R})$. To the best of our numerics, we find that this function is given by the holographic effective coupling $a_\mathrm{eff}=A_z|_{\rho=1}$,
\begin{equation}
\label{eq:holintegratedcurrent}
 J_y = \frac{\mu}{2\pi^2}\left(a_{\mathrm{eff},L}-a_{\mathrm{eff},R}\right) =  \mu \left(\sigma_{\mathrm{AHE},L}-\sigma_{\mathrm{AHE},R}\right)\,,
\end{equation} 
where we used Eq.~\eqref{eq:AHC_hol} and again identified $\alpha=1/16\pi^2$. From this result it is reasonable to expect the scaling of the current at zero temperature to be the same as that of the AHE \cite{Landsteiner:2015pdh}. 

Our result can be discussed from two different perspectives. First, we can think of the surface current as being a result of an axial magnetic field $\mathbf{B}_A$ localized at the surface. At first it may seem not surprising that our result Eq.~\eqref{eq:holintegratedcurrent} can be obtained from Eq.~\eqref{eq:uvintegratedcurrent} in a similar fashion as the AHE by replacing $a\mapsto a_{\mathrm{eff}}$. However this is a non-trivial result since Eq.~\eqref{eq:uvintegratedcurrent} is expected to be only valid in the homogeneous case. %Recall that $a_{\mathrm{eff}}$, being a bulk property of the phases, only incorporates the effect of a finite $M$. 
This implies that the higher order correction terms to Eq.~\eqref{eq:cme} due to an inhomogeneous field drop out during the integration. This is not obvious since in principle the higher order transport coefficients are not restricted in a way such that only total derivative terms appear in the current density.

Alternatively, one can interpret Eq.~\eqref{eq:holintegratedcurrent} using the Fermi arc picture. At low temperatures, it is meaningful to interpret the effective coupling $a_\mathrm{eff}$ as an observable of the \textit{effective} Weyl node separation in the system. Let us first consider the case of a non-trivial left and a trivial right side $a_\mathrm{eff,R}=0$, i.e.~the surface of a WSM. As one increases $a_\mathrm{eff,L}$, the Weyl nodes are further separated. Thus, the Fermi arc stretches, more surface states become available, and a higher current is generated. More precisely, the surface current is directly proportional to the Weyl node distance $a_\mathrm{eff,L}$. The interface between two WSMs is topologically equivalent to having a small trivial phase in between. Thus, two sets of surface states appear which, due to their chirality, are counter-propagating. Hence, the integrated current in Eq.~\eqref{eq:holintegratedcurrent} is given as the difference of the individual contributions of the two sides.

\section{Conclusions}
In this letter we have investigated the surface states of a holographic Weyl semimetal. This is a step further in the understanding of topological phases within strongly coupled quantum field theories using holography.

At low temperatures, a current is generated in the presence of a chemical potential along the surface of a holographic Weyl semimetal. Crucially, this current vanishes at the interface of two topologically trivial phases. Moreover, the integrated current only depends on the global order parameters, namely the anomalous Hall conductivities of the two phases. This is a strong indication of the topological nature of the surface states.

Furthermore, the dependence of the integrated current on the order parameters is completely determined by the anomaly. This allows for an interpretation in terms of Fermi arcs. 

This is a non-trivial statement since the local current is sensitive to higher order corrections in the effective inhomogeneous axial magnetic field naturally arising in a (realistic) interface. It would be interesting to investigate how our main result constrains the form of higher order transport coefficients in presence of inhomogeneous axial magnetic fields.

\section{Acknowledgements}
We thank Karl Landsteiner for insightful conversations. S.\,M.~acknowledges financial support by Deutsche Forschungsgemeinschaft
(DFG) GRK 1523/2. M.\,H.~acknowledges partial financial support by Deutsche Forschungsgemeinschaft (DFG) SFB 183. This paper is dedicated to Marcus Ansorg, a colleague and friend.

\appendix
\section{Apendix}

\subsection{Equations}

As described in the main text we used the static AdS Schwarzschild metric and the following ansatz for the fields:
\begin{align}
 \dd s^2 &= \frac{1}{\rho^2} \left( - f(\rho) \dd t^2 + \frac{\dd \rho^2}{f(\rho)}  + \dd x^2 + \dd y^2 + \dd z^2\right), \\
 A &= A_z(x,\rho) \dd z, \\
 V &= V_t(x,\rho) \dd t + V_y(x,\rho) \dd y, \\
 \Phi &= \phi(x,\rho).
\end{align}
Here, $f(\rho)=1-\rho^4$ is the blackening factor. We are then left with four equations for the field components:
\begin{align}
0 &= -\phi  \left(q^2 \rho ^2 A_z^2+m^2\right)+\left(\rho ^2 f'-3 \rho  f\right) \phi ^{(0,1)} \nonumber\\ 
  & \qquad +\rho ^2 f \phi ^{(0,2)}+\rho ^2 \phi ^{(2,0)} ,\\
0 &= -8 \alpha  \rho ^2 f A_z^{(1,0)} V_y^{(0,1)}+8 \alpha  \rho ^2 f A_z^{(0,1)} V_y^{(1,0)} \nonumber\\ 
  & \qquad -\rho  f V_t^{(0,2)}+f V_t^{(0,1)}-\rho  V_t^{(2,0)}, \\
0 &= -\rho  f V_y^{(0,2)}-\rho  V_y^{(2,0)}  -8 \alpha  \rho ^2 A_z^{(1,0)} V_t^{(0,1)} \nonumber\\ 
  & \qquad +8 \alpha  \rho ^2 A_z^{(0,1)} V_t^{(1,0)} +\left(f-\rho  f'\right) V_y^{(0,1)},\\
0 &= \left(\rho ^2 f'-\rho  f\right) A_z^{(0,1)}+\rho ^2 f A_z^{(0,2)} \nonumber\\ 
  & \qquad -2 q^2 \phi ^2 A_z +\rho ^2 A_z^{(2,0)}+8 \alpha  \rho ^3 V_t^{(1,0)} V_y^{(0,1)}\nonumber\\ 
  & \qquad-8 \alpha  \rho ^3 V_t^{(0,1)} V_y^{(1,0)}.
\end{align}
Partial derivatives are denoted by $F^{(k,l)}=\frac{\partial^{k+l}}{\partial x^k \partial \rho^l} F$. 
\subsection{Parameters}
We studied the dependence of the integrated current on the parameters $\alpha$, $q$, $\mu$, $l$ and found that $J_y$ is proportional to $\alpha \mu$ and independent of $l$. Moreover, the integrated current is also independent of $q$ as long as $q\geq 1$, see also Ref.~\cite{Copetti:2016ewq} for a recent discussion on the dependence of the phase transition on the parameters. We choose $q=1$, $\mu=0.1$, $l=4$. The scalar field mass is chosen to be $m^2=-3$ in order to agree with the dimension of the operator appearing in Eq.~(3) of the main text.   
\subsection{Numerical techniques}
%************************************************
We solved the system of partial differential equations numerically by means of pseudospectral methods. More specifically, we expanded the unknown functions in a Chebyshev polynomial basis set and solved for the collocation points on a Gauss-Lobatto grid. Concretely, we used the Newton-Raphson algorithm combined with the stabilised biconjugate gradient method (BiCGSTAB) as linear solver. Given the complexity of the problem, it is crucial to implement an appropriate domain decomposition. Domain decomposition techniques are widely used in modern approaches to partial differential equations. The idea is to split the domain of the PDEs into multiple subdomains,  $\Omega = \bigcup_i \Omega_i$. This provides many advantages and possibilities, among others the parallelisation of the numerics, the use of an adaptively refinable mesh and different coordinate systems on the subdomains.

\begin{figure}
\centering
 \begin{tikzpicture}[scale=0.7]
  \filldraw[lightgray] (-2,0) rectangle (2,3);
  
  \draw[<->] (-5,0)  -- (5,0)  node [right] {$\rho=1$};
  \draw[<->] (-5,3)  -- (5,3) node [right] {$\rho=0$};

  \draw[->] (-1,-0.5) -- (1,-0.5) node [midway, below] {$x$};
  \draw[<-] (-5,0.5) -- (-5,2.5) node [midway, left] {$\rho$};

  \draw[-] (0,3) -- (0,3.2) node [above] {$0$};
  \draw[-] (1,3) -- (1,3.2) node [above] {$l_1$};
  \draw[-] (2,3) -- (2,3.2) node [above] {$l_2$};
  \draw[-] (3.5,3) -- (3.5,3.2) node [above] {$l_3$};
  \draw[-] (-1,3) -- (-1,3.2) node [above] {$-l_1$};
  \draw[-] (-2,3) -- (-2,3.2) node [above] {$-l_2$};
  \draw[-] (-3.5,3) -- (-3.5,3.2) node [above] {$-l_3$};
  
  \draw[dashed] (0,3) -- (0,0);
  \draw[dashed] (1,3) -- (1,0);
  \draw[dashed] (2,3) -- (2,0);
  \draw[dashed] (3.5,3) -- (3.5,0);
  \draw[dashed] (-1,3) -- (-1,0);
  \draw[dashed] (-2,3) -- (-2,0);
  \draw[dashed] (-3.5,3) -- (-3.5,0);
  
  \node at (-4.3,1.5) {$\Omega_1$};
  \node at (-2.75,1.5) {$\Omega_2$};
  \node at (-1.5,1.5) {$\Omega_3$};
  \node at (-0.5,1.5) {$\Omega_4$};
  \node at (0.5,1.5) {$\Omega_5$};
  \node at (1.5,1.5) {$\Omega_6$};
  \node at (2.75,1.5) {$\Omega_7$};
  \node at (4.3,1.5) {$\Omega_8$};
 \end{tikzpicture}
\caption{Sketch of the chosen domain decomposition. The grey area represents the part of spacetime where a non-constant boundary value $a(x)$ is imposed. The domains $\Omega_1$ and $\Omega_8$ are non-compact and extend to $\pm\infty$ in $x$ direction.}
\label{fig:app:subdomains}
\end{figure}
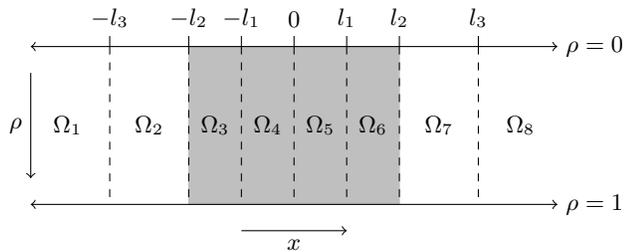

In our case we chose to decompose the two-dimensional domain into non-overlapping subdomains given by stripes $\Omega_i= I_i \times [0,1]$ where $I_i$ are suitable intervals, see Fig.~\ref{fig:app:subdomains}. Hence, the common boundaries are given by line segments $x=\text{const}$. The boundary value problem (BVP) on each subdomain $\Omega_i$ consists of the original BVP restricted to $\Omega_i$ supplied by internal boundary conditions of Poincar\'e-Steklov type \cite{mathew2008domain}.

The decomposition shown in Fig.~\ref{fig:app:subdomains} is justified as follows. The Gauss-Lobatto grid has the property that its density of points is the highest at the boundary. Dividing a domain at $x=\text{const}$ creates a new boundary and effectively increases the density of points there. Recall that the boundary profile $a(x)$ of the axial gauge field $A_z$ is chosen such that it interpolates on the interval $[-l,l]$ between the constant values $a_{L/R}$.  We chose to divide this interval into four subintervals, since we expect large gradients of the solution around $x=0$ and large gradients of the derivatives around $x=\pm l/2$. Next to the central subdomains we introduced ``buffer'' domains which separate the main from the asymptotic region. The outer domains extend to infinity in $x$ direction and have to be compactified to $[-1,1]\times[0,1]$ by a suitable coordinate transformation which we choose to be \footnote{Note that this choice of compactification corresponds effectively to the use of rational Chebyshev polynomials as a basis set.}:
\begin{align}
 x \mapsto & -1 - \frac{2}{x + l_3 - 1} & \text{for }x \in I_1, \\
 x \mapsto & 1 - \frac{2}{x - l_3 + 1} & \text{for }x \in I_8.
\end{align}
In all other cases, the $x$ coordinate is simply rescaled to $[-1,1]$. The treatment of the $\rho$ coordinate is described in next section. Finally, the parameters $l_1,l_2,l_3$ which determine the position of the domain boundaries are chosen as follows:
\begin{equation}
 l_1 = \frac{l}{2}, \quad l_2 = l, \quad l = 4, \quad l_3 = 15.
\end{equation}

%************************************************
\subsection{Non-analyticity and logarithms}\label{app:numerics:logarithms}
%************************************************

The boundary expansion of the fields contains logarithmic terms which effectively decrease the convergence rate of the Chebyshev expansion. Given the boundary conditions in Eqs.~(9-10) of the main text we find the following expansions:
\begin{align}
 V_t(x,\rho) &= \mu +  V_t^{(1)}(x) \rho^2 + \mathcal{O}(\rho^4),\\ 
 V_y(x,\rho) &= V_y^{(1)}(x) \rho^2  + \mathcal{O}(\rho^4), \\
 A_z(x,\rho) &= a(x) +\left( A_z^{(1)}(x) + \frac 1 2 \left(2 q^2 a(x) M^2 \right.\right.
  \nonumber\\	     & \qquad  - a''(x) ) \log \rho \bigg)\rho^2 + \mathcal{O}(\rho^4,\rho^4\log\rho), \\
 \varphi(x,\rho) &= M + \left( \varphi^{(1)}(x) + \frac 1 2 q^2 a(x)^2 M \log \rho  \right) \rho^2 \nonumber\\
	     & \qquad + \mathcal{O}(\rho^4,\rho^4\log\rho).
\end{align}

The lower the order where the first logarithmic term appears, the worse the convergence rate. However, there are strategies to enhance convergence in presence of logarithms. Here we use a simple ansatz involving a coordinate transformation:
\begin{equation}
 \rho = \frac{1}{2}\left(1+\sin\left(\frac \pi 2 \tilde{\zeta}\right)\right), \quad \text{with }\tilde{\zeta}\in[-1,1].
\end{equation}
This effectively shifts logarithmic singularities of order $k$ at the end points to order $2k$ and increases thus the convergence rate \cite{boyd1989asymptotic}.

However, there are further problems caused by the logarithms if we want to compute the non-normalisable modes in terms of $\rho$ derivatives. Namely, some $\rho$ derivatives are singular at $\rho=0$. Furthermore, the introduced sine transform has a singular Jacobian $J$ at the conformal boundary. Hence, we can not compute the $\rho$ derivatives using $D_\rho = J \cdot D_{\tilde{\zeta}}$. We solved these problems by a post-procession of the solutions as follows:
\begin{enumerate}
 \item The analytically known logarithmic terms are subtracted from the solutions $u$, which results in fields $\hat{u} = u - (\text{logarithms})$ with well-behaved $\rho$ derivatives at the conformal boundary.
 \item We interpolate the solutions on the $\tilde{\zeta}$ grid to a Chebyshev-Lobatto grid of the coordinate 
\begin{equation}
 \zeta = \sin\left(\frac \pi 2 \tilde{\zeta}\right) \in [-1,1],
\end{equation}
followed by the computation of $\rho$ derivatives using that $D_\rho = 2 D_\zeta$.
\end{enumerate}

%************************************************
\subsection{Numerical quality and convergence studies}\label{app:numerics:convergence}
%************************************************

Given a reference solution $u^\mathrm{ref}$, computed ideally on high-resolution grid, we use the following quantity as a measure for the convergence of a solution $u$:
\begin{equation}
\label{eq:app:conv_measure}
 \sup_{(x,\rho)\in\Omega} \left| u^\mathrm{ref}(x,\rho) - u(x,\rho) \right|.
\end{equation}
We compute the supremum as follows: We interpolate the piecewise-defined solutions $u^\mathrm{ref}$ and $u$ on a high-resolution equidistant grid of the whole $(x,\rho)$ domain. Using these interpolated solutions, we compute the maximal deviation as in Eq.~\eqref{eq:app:conv_measure}.

\begin{figure}
 \includegraphics[width=\linewidth]{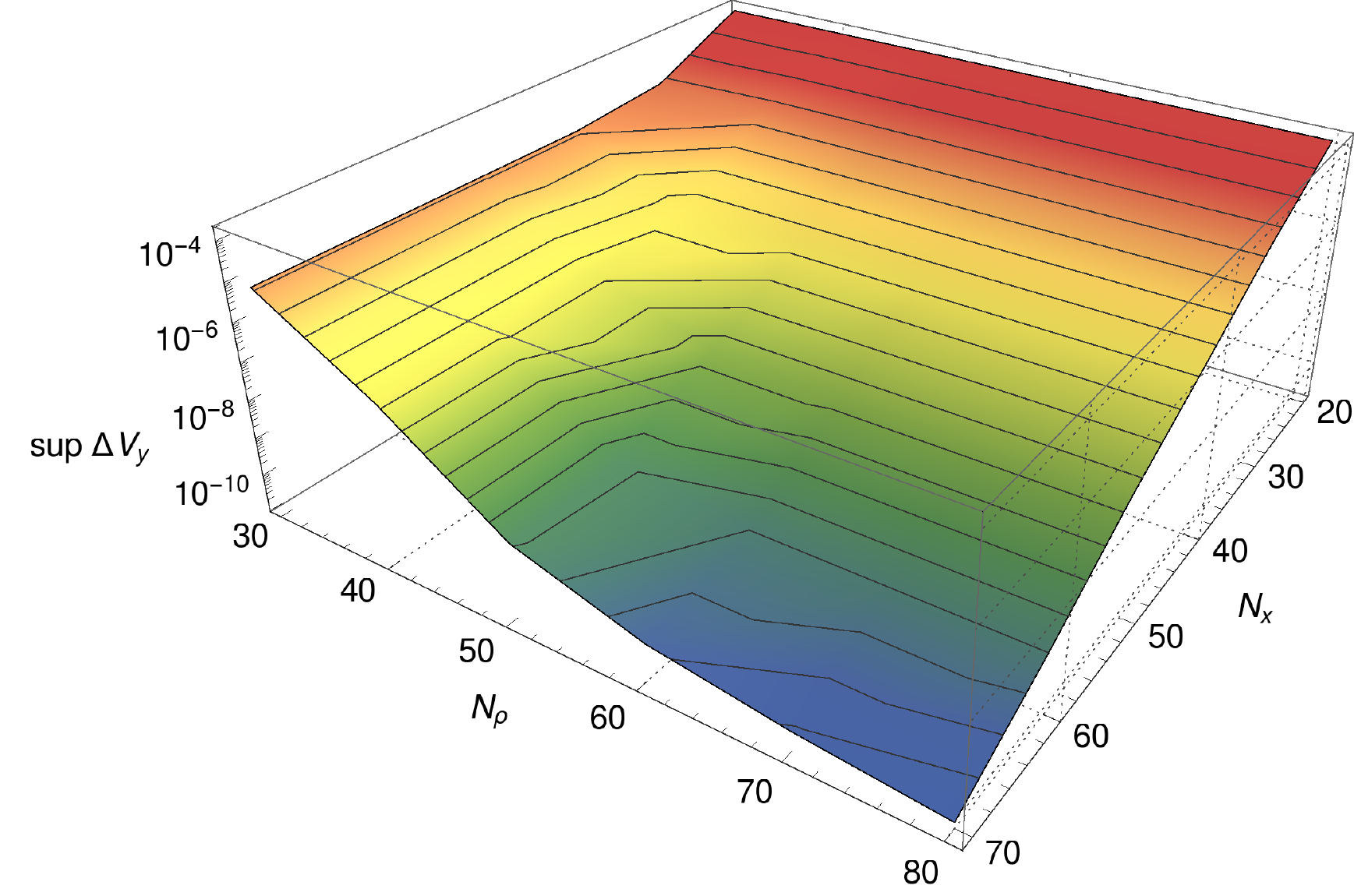}
 \caption{Maximal deviation in the field amplitude of $V_y$ as a function of the grid points in $x$ and $\rho$ direction. Note that this is not the total grid size but the grid size on each of the compact domains $\Omega_2$ to $\Omega_7$.}
 \label{fig:app:conv_plot}
\end{figure}

Figure \ref{fig:app:conv_plot} shows the convergence measure for $u=V_y$ as a function of the grid points. We varied the subgrid size $N_x \times N_\rho$ on each of the subdomains $\Omega_2$ to $\Omega_7$. As a reference solution, we used one which was computed on subgrids of size $100\times 80$. We can observe that the deviation decays exponentially with $N_x$. This is in line with spectral convergence theory since the spectral coefficients should decay exponentially with their order and hence also the overall correction to the solution. However, in $\rho$ direction, we see a deviation from this behaviour for $N_\rho \gtrsim 50$. This is due to the presence of logarithmic singularities at the conformal boundary which alter the asymptotic convergence. Without the sine transform, we would have seen this behaviour already for smaller $N_\rho$.

We used $N_\rho=55$ and $N_x=35$ since this was sufficient to achieve the desired accuracy.

From the convergence studies, we can estimate that the relative numerical error of the solution is less than $10^{-5}$. To get an error on the integrated current $J_y$, we estimate that differentiation and integration could increase the error by about two magnitudes to $10^{-3}$. We can test this by checking how much $J_y$ deviates from zero for $a_L=a_R$. We found that the higher $a_L=a_R$, the higher is the deviation. However, we found that it is at most $10^{-5}$ and hence much less than the error estimate.

\bibliographystyle{apsrev4-1} 
\bibliography{Sb}
%apsrev4-1

\end{document}